\documentclass[twocolumn,amsmath,amssymb]{revtex4-2}
\usepackage{graphicx}
\usepackage{multirow}
\usepackage{subfigure}
\usepackage{epsfig}
\usepackage{color}
\usepackage{graphics}
\usepackage{epstopdf}
\usepackage{adjustbox}
\usepackage{float}
\usepackage{latexsym}
\usepackage{bm}
\usepackage{lipsum}
\usepackage{datetime}
\usepackage{wasysym}
\usepackage{slashed}
\usepackage{placeins}
\usepackage{cancel}
\usepackage{array}
\makeatletter \renewcommand{\@dotsep}{10000} \makeatother
\usepackage{indentfirst}
\usepackage{natbib,hyperref}
\usepackage[justification=raggedright, margin=5mm,font=sf,small,labelfont=bf, format=plain]{caption}
\def\lsim{\raise0.3ex\hbox{$\;<$\kern-0.75em\raise-1.1ex\hbox{$\sim\;$}}}
\def\gsim{\raise0.3ex\hbox{$\;>$\kern-0.75em\raise-1.1ex\hbox{$\sim\;$}}}

\newcommand{\be}{\begin{equation}}
\newcommand{\ee}{\end{equation}}
\def\bea{\begin{eqnarray}}
\def\eea{\end{eqnarray}}

\mathchardef\mhyphen="2D

\newcommand{\beq}{\begin{equation}}
	\newcommand{\eeq}{\end{equation}}

\newcommand{\spheno}{\textsc{SPheno 4.0.4}}

\usepackage[nodisplayskipstretch]{setspace}
\def\be{\begin{equation}}
	\def\ee{\end{equation}}
\def\bea{\begin{eqnarray}}
	\def\eea{\end{eqnarray}}

\begin{document}
\title{Light $Z'$ Signatures at the LHC}
\author{
Ya\c{s}ar Hi\c{c}y\i lmaz$^{1,2}$\footnote{E-mail: Y.Hicyilmaz@soton.ac.uk},
Shaaban Khalil$^{3}$\footnote{E-mail: skhalil@zewailcity.edu.eg}  and Stefano Moretti$^{2,4}$\footnote{E-mail: S.Moretti@soton.ac.uk; Stefano.Moretti@physics.uu.se} }

\affiliation{
$^1$Department of Physics, Bal\i kesir University, TR10145, Bal\i kesir, Turkey \\
$^2$School of Physics and Astronomy, University of Southampton, Highfield, Southampton SO17 1BJ, United Kingdom\\
$^3$Center for Fundamental Physics, Zewail City of Science and Technology, 6 October City, Giza 12588, Egypt\\
$^4$Department of Physics \& Astronomy, Uppsala University, Box 516, SE-751 20 Uppsala, Sweden}
\date{\today}

\begin{abstract}
{In this work, we discuss a distinctive $pp\to {\rm Higgs}\to Z'Z'\to 4l$ ($l=e,\mu$) signal at the Large Hadron Collider (LHC), where the `Higgs' label refers to the SM-like Higgs state discovered in 2012 or a lighter one in the framework of a theoretical model embedding a spontaneously broken $U(1)'$ symmetry in addition to the  Standard Model (SM) gauge group. The additional $U(1)'$ symmetry generates a very light $Z'$ state, with both vector and axial (non-universal) couplings to fermions, which are able to explain the so-called Atomki anomaly, compliant with current measurements of the Anomalous Magnetic Moments (AMMs) of electron and muon as well as beam dump experiments. We show that the cross section for this process should be sufficiently large to afford one with significant sensitivity during Run 3 of the LHC.}

\end{abstract}
\maketitle

\section{Introduction}
\label{sec:intro}
A light neutral $Z'$ boson (often dubbed a `dark photon'), with mass of order 17 MeV, provides a natural explanation for the clear anomaly observed by the Atomki collaboration \cite{Gulyas:2015mia} in the decay of excited states of Beryllium \cite{Krasznahorkay:2015iga,Sas:2022pgm,Krasznahorkay:2017gwn,Krasznahorkay:2017bwh,Krasznahorkay:2017qfd,Krasznahorkay:2018snd}. Furthermore, several studies have been conducted to investigate the effects of such light $Z'$ on the AMM of the electron ($a_e$) and muon ($a_\mu$)  as well as  $B$ anomalies such as $R_{K^{(*)}}$  \cite{Barman:2021yaz,Bodas:2021fsy,Fayet:2020bmb,Nomura:2020kcw,Seto:2020jal,Hati:2020fzp,Pulice:2019xel,DelleRose:2017xil}. 

In this letter we analyse some LHC signatures of a light $Z'$ associated with a non-universal $U(1)'$ extension of the Standard Model (SM). This type of scenario has been shown to account for both the  Atomki anomaly and $a_{e,\mu}$ results  \cite{DelleRose:2018eic}.  In addition, we revisit the contributions of such light $Z'$ to these observables to see how the most recent experimental results constrain the associated couplings.

We focus on a non-universal $U(1)'$ extension of the SM in which the kinetic term in the Lagrangian is given by
\begin{equation}
	\label{eq:KineticL}
	\mathcal{L}_\mathrm{kin} = - \frac{1}{4} \hat F_{\mu\nu} \hat F^{\mu\nu} - \frac{1}{4} \hat F'_{\mu\nu} \hat F^{'\mu\nu} - \frac{\eta}{2} \hat F'_{\mu\nu} \hat F^{\mu\nu},
\end{equation}
where $\eta$ quantifies the mixing between the SM $U(1)_Y$ and  extra $U(1)'$. After the diagonalization of Eq.~(\ref{eq:KineticL}), the covariant derivative can be written as
\begin{equation}
\label{CovDer}
	{\cal D}_\mu = \partial_\mu + .... + i g_1 Y B_\mu + i (\tilde{g} Y + g' z) B'_\mu, 
\end{equation}
where $Y$ and $g_1$ are the hypercharge and its gauge coupling while $z$ and $g'$ are the $U(1)'$ charge and its gauge coupling. Further, $\tilde{g}$ is the mixed gauge coupling between the two groups. The $U(1)'$ symmetry is broken by a new SM singlet scalar, $ \chi $, with $U(1)'$ charge $z_\chi$ and  Vacuum Expectation Value (VEV) $v'$. The scalar potential for the  Higgs fields can be written as 
\begin{equation}
	\label{eq:HM}
\!\!V(H,\chi)\!=\!-\mu^2|H|^2 \!+\!\lambda |H|^4 \!-\!\mu_\chi^2 |\chi|^2 \!+\! \lambda_\chi |\chi|^4 \!+\! \kappa |\chi|^2|H|^2.
\end{equation}
Here, $H$ is the SM Higgs doublet while $\kappa$ is the mixing parameter which connects that SM and $ \chi $ Higgs fields. After Electro-Weak Symmetry Breaking (EWSB), for $  \mu^{2} =  \lambda v^{2}  + \frac{1}{2} \kappa v'^{2}$ and $ \mu_{\chi}^{2} =  \lambda_{\chi} v'^{2}  + \frac{1}{2} \kappa v^{2}$, the Higgs mass matrix in the $  (h_2, h_1) $ basis can be written as
\begin{equation} 
m^2_{h_2 h_1} = \left( 
\begin{array}{cc}
2 \lambda v^{2}  & \kappa v v' \\ 
 \kappa v v'  &2 \lambda_{\chi} v'^{2} \end{array} 
\right), 
\end{equation} 
where $ h_2 $ is dominantly the SM-like Higgs boson while the exotic state $ h_1 $ is dominantly the singlet Higgs ($ \chi $-like). In this work, we consider $ m_{h_1}  < m_{h_2} $ and the $h_1 \to Z' Z'$ branching ratio $\geq 0.95$, which are be compatible with experimental results. The SM-like Higgs boson $ h_2 $ can decay to  $ Z' $ pairs too, proportionally to $ \kappa $. Moreover, the spontaneous breaking of the $U(1)'$ symmetry implies the existence of a mass term $m_{Z'}=g' z_\chi v'$. Thus, if $g' \sim {\cal O}(10^{-4} - 10^{-5})$, $M_Z'$ would be of order ${\cal O}(10)$ MeV. 
It is worth noting that we adopt non-universal   charge assignments of the SM particles under $U(1)'$, as discussed in Ref.~\cite{DelleRose:2018eic}. These assignments satisfy  anomaly cancellation conditions,  enforcing a gauge invariant Yukawa sector of the third fermionic generation and family universality in the first two while  not allowing coupling between $Z'$ and light neutrinos. 

The Neutral Current (NC) interactions of this additional vector boson with the SM fermions are given as
\begin{equation}
	\label{eq:NeuCurLag}
	\mathcal{L}_\mathrm{NC}^\mathrm{Z'}  = -\sum_f \bar \psi_f \gamma^\mu \left( C_{f, L} P_L + C_{f, R} P_R \right) \psi_f Z'_\mu,
\end{equation}
where Left (L) and Right (R) handed coefficients are written as 
{\small \begin{eqnarray}
	C_{f,L} \!&\!=\!&\!  - g_Z \sin \theta' \left( T^3_f - \sin^2 \theta _W Q_f \right) \!+\! ( \tilde g Y_{f, L} \!+\! g' z_{f, L})  \cos \theta'\!,~~~ \nonumber\\
	C_{f,R} &=&  g_Z \sin^2 (\theta _W) \sin(\theta') Q_f + ( \tilde g Y_{f, R} + g' z_{f, R}) \, \cos (\theta ').
	\label{couplings}
\end{eqnarray}}
The parameters given in these expressions can be found in Ref.~\cite{DelleRose:2018eic}.

The contribution of this $ Z' $ gauge boson to the AMMs of the charged leptons $a_f$, for $ f= e, \mu, \tau$  is given by \cite{Leveille:1977rc}
\begin{eqnarray}
\Delta a_\textit{f} &=& \frac{m_\textit{f}^2}{4\pi^2 m_{Z'}^2} \Big( C_{\textit{f}, V}^2 \int_0^1 \frac{x^2 (1-x)}{1 - x + x^2 m_\alpha^2/m_{Z'}^2} dx \nonumber\\
&-& C_{\textit{f}, A}^2 \int_0^1 \frac{x (1-x) (4 - x) + 2 x^3 m_\textit{f}^2/ m_{Z'}^2}{1 - x + x^2 m_\textit{f}^2/m_{Z'}^2} dx \Big), .
\label{eq:gm2sfull}
\end{eqnarray}
where $ C_{f, V} = \frac{C_{f,R} + C_{f,L}}{2} $ and $ C_{f, A} = \frac{C_{f,R} - C_{f,L}}{2} $. For the limits $m_f \ll m_{Z'}$ and $m_f \gg m_{Z'}$, Eq. (\ref{eq:gm2sfull}) reduces to \cite{Bodas:2021fsy}
\begin{equation}
	\Delta a_f \simeq \begin{cases}
		m_f^2 \left( C_{f, V}^2 - 5 C_{f, A}^2 \right)/(12\pi^2 m_{Z'}^2)~, & m_f \ll m_{Z'} \, , \\
		(m_{Z'}^2 C_{f, V}^2 - 2m_f^2 C_{f, A}^2)/(8\pi^2 m_{Z'}^2) ~, & m_f \gg m_{Z'} \, .
	\end{cases}
	\label{gm2s}
\end{equation}
 
It is important to note that the contribution of the $ Z' $ to the  AMMs of leptons is primarily determined by their vector and axial couplings which are expressed in Eq.~(\ref{couplings}), as well as the mass of the $ Z' $ boson. Using the charge assignments in Ref.~\cite{DelleRose:2018eic}, one can find the contributions to AMMs of electron and muon as 
 \begin{eqnarray}
\Delta a_{e} &=&  -3.6x10^{-6} {g'}^2 + 6.5x10^{-6} g^\prime \tilde{g} + 4.6x10^{-6} \tilde{g}^2 \, , \nonumber\\
\Delta a_{\mu} &=&  -0.99 {g'}^2 + 0.00775 g^\prime \tilde{g} + 0.00543 \tilde{g}^2 \, .
\label{eq:deltagm2}
\end{eqnarray}

Furthermore, the vector and axial couplings of the quarks are important in explaining the Atomki anomaly via the transition.${}^8 \textrm{Be}^* \rightarrow {}^8 \textrm{Be} \, Z'$ \cite{Kozaczuk:2016nma}.  In particular, the contribution of the quark axial couplings $ C_{q, A} $ in this transition is greater than that of the vector couplings $ C_{q, V} $ because the $ C_{q, A} $ and $ C_{q, V} $ terms are proportional to $k/M_{Z'}$ and $k^3/M_{Z'}^3$ (where $k$ is the small momentum of the $ Z' $), respectively \cite{Feng:2016jff}. According to $ U(1)' $ charges in the model, $ |C_{q, A}| $ equals to $ g' $.

\section{Computational setup and experimental constraints}
\label{sec:comp}
\begin{figure}[h]
	\centering
	\includegraphics[scale=0.48]{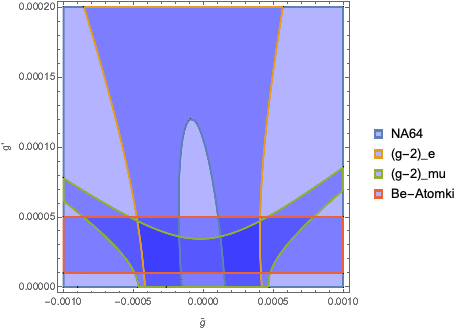}
	\caption{ Allowed parameter space mapped on the $(g',\tilde{g})$ plane for $Z'$ mass of 17 MeV against four different experimental constraints.}
	\label{fig:new_bound}
\end{figure}

In our numerical analysis, we have employed  $\spheno$ \cite{Porod:2003um,Porod:2011nf,Braathen:2017izn} generated with  SARAH 4.14.3 \cite{Staub:2013tta,Staub:2015kfa}. In Fig.~\ref{fig:new_bound}, we show the portion of ($ g' $, $ \tilde{g} $) parameter space that satisfies the current experimental bounds from $ (g-2)_{e,\mu} $,  the ${}^8 \textrm{Be}^{*}$ anomaly and  NA64 (as well as electron beam dump experiments) \cite{Davoudiasl:2018fbb,Morel:2020dww,Muong-2:2021ojo,NA64:2019auh}. Here, the darkest shaded blue regions comply with  all such constraints. {Considering the similar plot in Ref.~\cite{DelleRose:2018eic}, one can see that the allowed regions have changed slightly}. During the scanning of the $U(1)'$ parameter space, within the ranges specified in Tab.~\ref{paramSP}, the Metropolis-Hastings algorithm has been used. After data collection, we implement Higgs boson mass bounds \cite{CMS:2012qbp,ATLAS:2012yve} as well as constraints from Branching Ratios (BRs) of $B$ decays such as $ {\rm BR}(B \rightarrow X_{s} \gamma) $ \cite{HFLAV:2012imy}, $ {\rm BR}(B_s \rightarrow \mu^+ \mu^-) $ \cite{LHCb:2012skj} and $ {\rm BR}(B_u\rightarrow\tau \nu_{\tau}) $ \cite{HFLAV:2010pgm}. We have also bounded the $ Z/Z' $ mixing to be less than a few times $ 10^{-3} $ as a result of EW Precision Tests (EWPTs) \cite{Erler:2009jh}.

\begin{table}[H]
	\begin{tabular}{c|c||c|c}
		Parameter  & Scanned range & Parameter      & Scanned range \\
		\hline
		$g'$ & $[10^{-5}, 5 \times 10^{-5}]$      & $\lambda$ & $[0.132, 0.125]$ \\
		$\tilde{g}$        & $[-10^{-3}, 10^{-3}]$ & ${\lambda}_{\chi}$ & $[10^{-5}, 10^{-3}]$ \\
		$v_S$ & $[0.1, 1]$ TeV  &  $\kappa$  & $[10^{-6}, 10^{-3}]$ \\
	\end{tabular}
	\caption{\sl\small Scanned parameter space of our model.}
	\label{paramSP}
\end{table}

The experimental constraints can be summarized as follows:
\begin{equation}
	\setstretch{1.8}
	\begin{array}{l}
		m_h  = 122-128~{\rm GeV}({\rm as~our~masses~are~lowest~order}),
		\\
		2.99 \times 10^{-4} \leq
		{\rm BR}(B \rightarrow X_{s} \gamma)
		\leq 3.87 \times 10^{-4} \; (2\sigma~{\rm tolerance}),
		\\
		0.15 \leq \dfrac{
			{\rm BR}(B_u\rightarrow\tau \nu_{\tau})}
		{{\rm BR}(B_u\rightarrow \tau \nu_{\tau})_{\rm SM}}
		\leq 2.41 \; (3\sigma~{\rm tolerance}),
		\\
		\Delta a_e = (4.8 \pm 9.0) \times 10^{-13} \; (3\sigma~{\rm tolerance}),
		\\
		\Delta a_{\mu} = (2.51 \pm 1.77) \times 10^{-9} \; (3\sigma~{\rm tolerance}).
		\label{constraints}
	\end{array}
\end{equation}

Additionally, the cross section values for the given processes at the LHC have been calculated by using {\textsc CalcHEP/Madgraph5} \cite{Belyaev:2012qa,Alwall:2014hca}.

\section{Results}
\label{sec:results}

\subsection{Constraints on Parameter Space}
\label{sec:param}

\begin{figure}[H]
	\centering
	\includegraphics[scale=0.48]{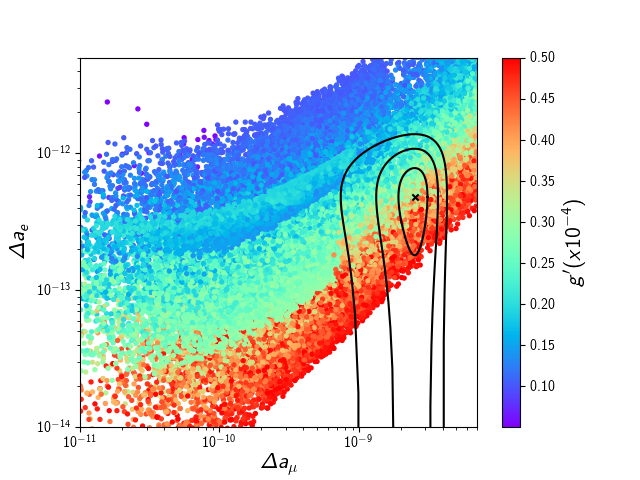}
	\includegraphics[scale=0.48]{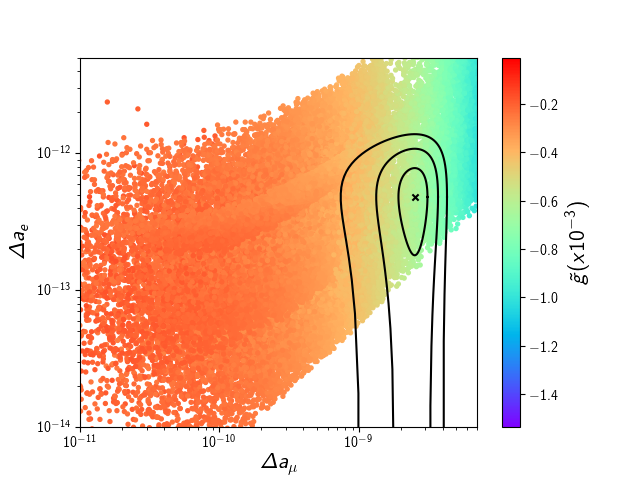}
	\caption{Results for $ g' $ (top) and $ \tilde{g} $ (bottom) in terms of $ (g-2)_e $ vs $ (g-2)_{\mu} $. Each solid line from inner to outer  represents 1$ \sigma $, 2$\sigma$ and 3$\sigma$ bounds from the experimental central values  in {Eq.~(\ref{constraints}).}}
	\label{fig:g2_gp_gtilde}
\end{figure}

{In this section, we will first present the dependence of $ \Delta a_{\mu} $ and $ \Delta a_{e} $ upon the fundamental parameters $ g' $ and $ \tilde{g} $.  
Fig.~\ref{fig:g2_gp_gtilde} depicts $ \Delta a_{\mu} $ vs $ \Delta a_{e} $ with different color bars that show $ g' $ (top panel) and $ \tilde{g} $ (bottom panel). Herein, considering Eq. (\ref{eq:deltagm2}), one can learn about the favored $ (g' , \tilde{g}) $ space in order to obtain AMMs for each 1$ \sigma $, 2$\sigma$ and 3$\sigma$ value. Additionally to Fig. \ref{fig:new_bound}, the panels in this figure give us significant information about how the different slices of  parameter
space are correlated to the AMMs. As seen from the plots, the experimental bounds of $ \Delta a_{\mu} $ and $ \Delta a_{e} $ within $3 \sigma$ allow for a narrow range in $ \tilde{g} $, namely, $ -0.6\times10^{-3} \lesssim \tilde{g} \lesssim -0.4\times 10^{-3} $ while $ g' $ lies in the range of $ 0.2\times10^{-4} \lesssim g' \lesssim 0.5\times 10^{-4} $. It is important to note that each area between the AMMs contours covers varied regions of the $ (g' , \tilde{g}) $ plane within these bounds.}

\begin{figure}[H]
	\centering
	\includegraphics[scale=0.48]{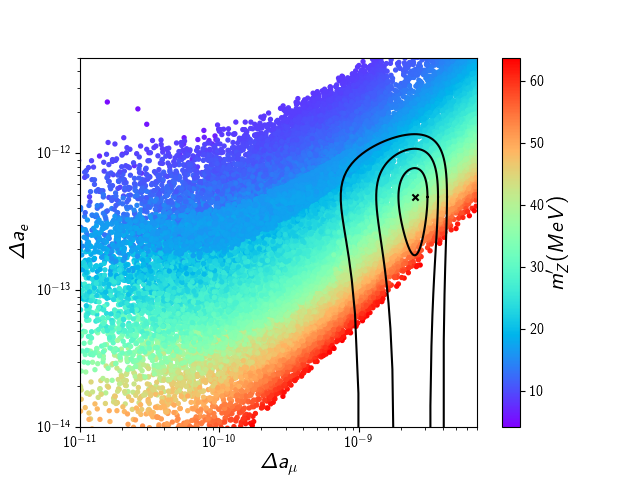}
	\includegraphics[scale=0.48]{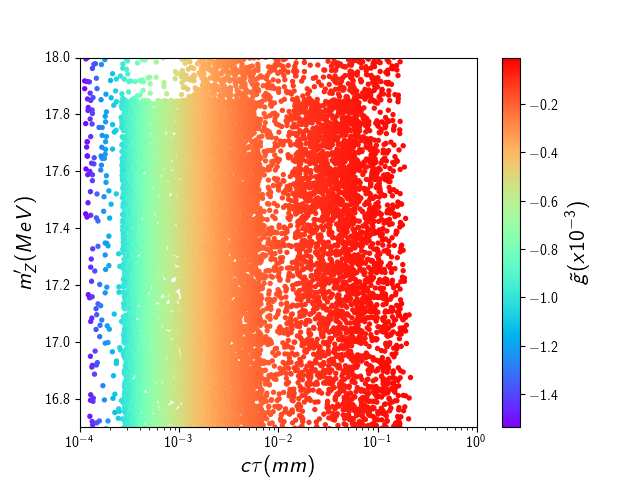}
	\caption{Results for $ m_{Z'} $   in terms of $ (g-2)_e $ vs $ (g-2)_{\mu} $ (top) and for $\tilde g$ in terms of $ m_{Z'} $  vs the proper lifetime of the $Z'$ (for 
		$m_Z' \approx 17$ MeV).}
	\label{fig:Ztw_mZp}
\end{figure}

{Now, let us focus on $ Z' $ properties, such as its mass $m_{Z'}$ and proper lifetime $ c\tau $. In the top panel of Fig.~\ref{fig:Ztw_mZp}, we demonstrate how $Z'$ mass solutions showed in the color bar correlate with  $ \Delta a_{\mu}$ and $  \Delta a_{e} $. Herein, our 1$\sigma $ solutions are excluded for $m_Z' \approx 17$ MeV, the best value satisfying the  Atomki anomaly.  This exclusion mainly arises from the tension between the AMMs and  $m_Z'$. As can be seen from Eq. (\ref{eq:deltagm2}), $ g' $ provides significant contributions to $ \Delta a_{\mu} $ and $ \Delta a_{e} $ while it  also impacts the $Z' $ mass since $m_{Z'}=g' z_\chi v'$. Therefore, such a $Z'$ mass value to fit the Atomki anomaly puts a limit on $ g' $ when it  is located out of 1$\sigma $ region as shown in the top panel of Fig.~\ref{fig:g2_gp_gtilde}. } We also examine the $Z'$ lifetime since it is crucial to explore potentially displaced signatures at the LHC. The plot at the bottom of Fig.~\ref{fig:Ztw_mZp} showcases the proper lifetime of $Z'$ in milimeters over the mass range  $ 16.7$  MeV $\lesssim m_Z' \lesssim 18$ MeV  while the color bar indicates $\tilde{g}$. As mentioned in Ref.~\cite{Lagouri:2022oxv}, for small values of $\lvert \tilde{g} \rvert $, the $Z'$ lifetime becomes longer. Considering the $ \tilde{g} $ solutions which fulfill all experimental conditions, the lifetime of the $Z'$ should be $ \sim 10^{-3}$ mm, which is not sufficient to produce a displaced detector signal. 

\subsection{$Z'$ production at the LHC}
\label{sec:ZpAtLHC}

Now, we will study the collider signatures of our light $Z'$ boson in three different channels at the LHC: Drell-Yan (DY) and $Z'$
pair production through both SM-like Higgs $ h_2 $ and  exotic Higgs $h_1$ mediation, wherein we consider both fully leptonic and semi-leptonic final states.

\subsubsection{Drell-Yan}
\label{sec:drellyan}

\begin{figure}[H]
	\centering
	\includegraphics[scale=0.48]{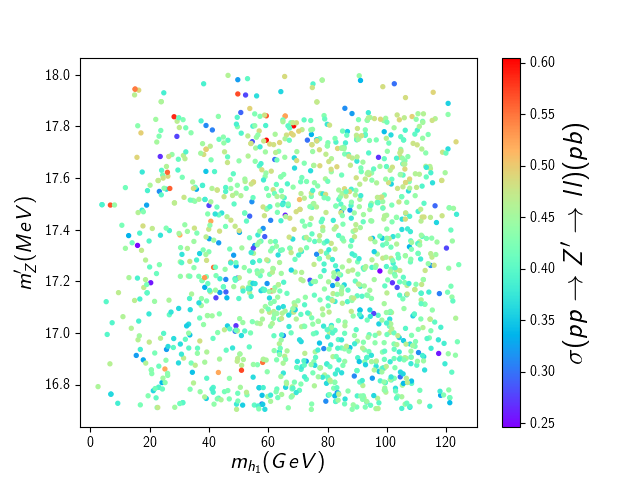}
	\caption{Results for  $\sigma(pp\to Z'\to ll)$ ($l=e,\mu$) in terms of $m_{h_1}$ vs $m_{Z'}$, for $ \sqrt{s}= 14 $ TeV.}
	\label{fig:Zptoll}
\end{figure}

At the LHC, the most favored process for a light $Z'$ boson is the DY channel, where it can  directly be generated via $ q\bar q $ fusion in $s$-channel. In Fig.~\ref{fig:Zptoll}, we present the dilepton production cross section via our light $Z'$ resonance. Although the
corresponding $Z'$  production and decay rates are always  large for $m_{Z'}\approx 17$ MeV, the process is difficult to detect given the very light $Z'$, implying very soft decay products. Hence, our $Z'$ is not really constrained by present LHC data, so that all points presented in this plot (at $ \sqrt{s}= 14 $ TeV) are amenable to experimental investigation during Run 3. However, a more striking signature would be $Z'$ pair production, to which we turn next.

\subsubsection{$Z'$ Pair Production via SM-like Higgs Mediation}
\label{sec:ZpSMH}

\begin{figure}[H]
	\centering
	\includegraphics[scale=0.48]{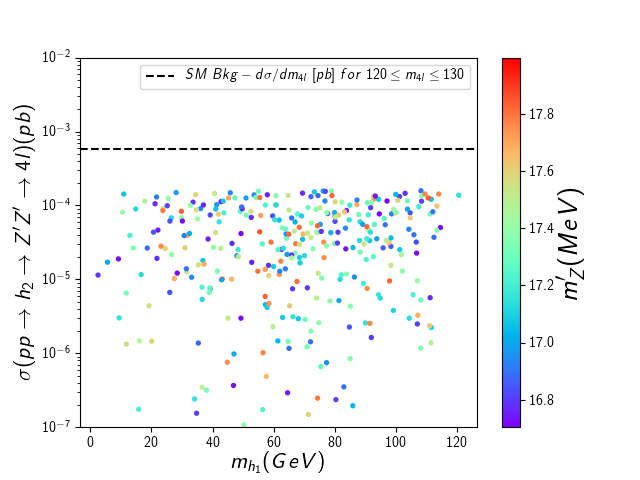}
	\caption{Results for $m_{Z'}$ in terms of  $m_{h_1}$ vs $\sigma (pp \to h_2 \to Z' Z' \to 4l)$, for $ \sqrt{s}= 14 $ TeV. The dashed line shows the SM cross section of $ pp  \to 4l $ for the mass region  of $ 120 \leq m_{4l} \leq 130 $.}
	\label{fig:h2toZp4lBRbounded}
\end{figure}

{As $ m_{Z'} \ll m_{h_{1,2}}/2 $, our light $Z'$ boson can be pair produced via both Higgs bosons $h_1$ and $h_2$.  Let us start with SM-like Higgs mediation. In Fig. \ref{fig:h2toZp4lBRbounded}, we present the cross section of the ensuing four-lepton final state at $ \sqrt{s}= 14 $ TeV for the solutions satisfy all experimental bounds considered so far, with the additional  requirement BR$(h_2\to Z'Z'\to 4l)<5\times10^{-6}$, following 
 ATLAS \cite{ATLAS:2021ldb} and CMS \cite{CMS:2021pcy} results. The color bar  shows the mass of the $Z'$ while the dashed line shows the SM cross section for $ pp  \to 4l $ , $ \sigma_{\rm SM} \approx 0.5 $ fb, for the mass region   $ 120$ GeV $ \leq m_{4l} \leq 130 $ GeV. As can be seen, the rates for $\sigma (pp \to h_2 \to Z' Z' \to 4l)$ can be rather large, up to $ \approx 0.1 $ fb, over a wide range of $m_{h_1}$, including very small values of the latter, which in turn call for studying $h_1$ mediation, in our next section. Considering the solutions with 0.1 fb cross sections without any cuts, in order to get an excess with $ 3\sigma $ significance ($ S/ \sqrt{B} $) in the mass region $m_{4l} \approx 125$ GeV  in Run 3, it is needed to gather data corresponding integrated luminosity of  500 fb$^{-1}$, at least.}

%

\subsubsection{$Z'$ Pair Production via  Exotic Higgs Mediation}
\label{sec:4l}

\begin{figure}[H]
	\centering
	\includegraphics[scale=0.48]{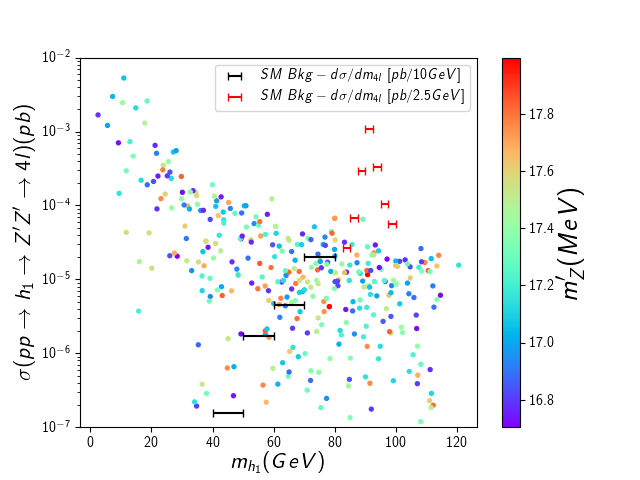}
	\caption{Results for $m_{Z'}$ in terms of $m_{h_1}$ vs $\sigma (pp \to h_1 \to Z' Z' \to 4l)$, for $ \sqrt{s}= 14 $ TeV. Black and red lines show the SM differential cross sections as a function of four lepton invariant mass with 10 GeV and 2.5 GeV bin size, respectively.}
	\label{fig:h1toZp4l}
\end{figure}

{In this  final part, we investigate  $Z'$ pair production via the new exotic Higgs, $h_1$.  Fig.~\ref{fig:h1toZp4l} shows  $\sigma (pp \to h_1 \to Z' Z' \to 4l)$  correlated to $m_{h_1}$ as well as $m_{Z'}$,  for the same parameter space considered in the previous plot (again, $\sqrt s=14$ TeV). In this case, the four-lepton rate can be larger than $10\times10^{-3}$ pb for a light ${h_1}$  while reaching  $2\times 10^{-5}$  pb for $m_{h_1}$ tending to $m_{h_2}$. Black and red lines show the SM differential cross sections for $\sqrt s=14$ TeV, calculated by MadGraph \cite{Stelzer:1994ta},  as a function of the four-lepton invariant mass with 10 GeV and 2.5 GeV bin size, respectively. We especially use 2.5 GeV bin size for the mass region  $ m_{h_1} \geq 80 $ GeV, where the SM background is dominant. Similar results for the SM differential cross sections for thye four-lepton final state at $\sqrt s=13$ TeV were published by the ATLAS Collaboration in Fig. 5 of Ref. \cite{ATLAS:2021kog}. As seen from the figure, for $ m_{h_1} \leq 85$ GeV, we have many solutions giving a clear signal around  $ m_{4l} \approx m_{h_1}$ in the four-lepton invariant mass distribution, due to a small SM background. The mass region $ 85$ GeV $ \leq m_{h_1} \leq 95 $ GeV is instead challenging since the $q\bar q \to Z \to 4l$ channel is  dominant. We also have  a small window in the mass region of $ 95$ GeV $ \leq m_{h_1} \leq 100$ GeV. Considering the solutions with largest cross sections, $ \sigma \approx 0.01$ fb without any cuts, it is possible to obtain an excess with 2.5 $ \sigma $ significance using  data corresponding to an integrated luminosity of  3000 fb$^{-1}$.  Herein, in order to reduce this background, it is possible to use invariant mass cuts for leading and$ / $or sub-leading lepton pairs. Hence, the $h_1$ mediated process, depending on the $m_{h_1}$ value, producing a $Z'$ pair decaying into four-lepton final states, can actually be  the best way to  access both the new Higgs and new gauge sectors of our scenario. }

%

\section{Conclusion}
\label{sec:conclusion}
In summary, a rather simple theoretical framework, assuming a non-universally coupled (to fermions) $Z'$ boson, with a mass of $O(10)$ MeV, emerging from a spontaneously broken $U(1)'$  group additional to the SM gauge symmetries, is able to explain several data anomalies currently existing at low energies while predicting a clear signal at high energies. Namely, the latter is a very clean process, potentially extractable at the upcoming Run 3 of the LHC, i.e., $pp\to h_i\to Z'Z'\to 4l$ ($l=e,\mu$), where $h_1$ and $h_2$ are the new Higgs state associated to the additional gauge group and the SM-like one already discovered, respectively. Hence, a new `golden channel' involving again four leptons in the final state could soon give access to both a new neutral Higgs and gauge boson.

\section*{Acknowledgements}
SK is partially supported by the Science,
Technology and Innovation Funding Authority (STDF) under Grant No. 37272. SM
is supported in part through the NExT Institute and the STFC Consolidated Grant No.
ST/L000296/1. The work of YH is supported by The Scientific and Technological Research Council of Turkey (TUBITAK) in the framework of the 2219-International Postdoctoral Research Fellowship Programme and by  Balikesir University Scientific Research Projects with Grant No. BAP-2022/083.

\end{document}